\documentclass[a4paper,fleqn]{cas-dc}

\usepackage[authoryear]{natbib}

\def\tsc#1{\csdef{#1}{\textsc{\lowercase{#1}}\xspace}}
\tsc{WGM}
\tsc{QE}

\begin{document}
\let\WriteBookmarks\relax
\def\floatpagepagefraction{1}
\def\textpagefraction{.001}

\shorttitle{The recent crossing of the 7:3 resonance between Ganymede and Callisto}    

\shortauthors{Giacomo Lari and Mattia Rossi}  

\title[mode=title]{The recent crossing of the 7:3 resonance between Ganymede and Callisto}

\author[1]{Giacomo Lari}[orcid=0000-0003-0868-8498]

\cormark[1]

\ead{giacomo.lari@unipi.it}

\affiliation[1]{organization={Department of Mathematics, University of Pisa},
            addressline={Largo Bruno Pontecorvo 5}, 
            city={Pisa},
            postcode={56127}, 
            country={Italy}}

\author[1]{Mattia Rossi}[orcid=0000-0003-2932-5688]

\ead{mattia.rossi@dm.unipi.it}

\cortext[1]{Corresponding author}

\begin{abstract}
Io, Europa and Ganymede are locked in the well-known Laplace resonance, while Callisto is not, even though it is very close to the 7:3 mean motion resonance with Ganymede. When considering the estimated tidal dissipation in the Jovian system and the consequent orbital migration rate of the moons, it follows that the 7:3 resonance should have been encountered around two million years ago. Through accurate numerical simulations, we investigated the dynamical effects of the 7:3 resonance crossing on the recent orbital evolution of the Galilean moons. We found that the most probable dynamical pathway is that Ganymede and Callisto were not captured into resonance, not even temporarily, while their eccentricities experienced a downward kick because of the resonance crossing, reducing their mean values of about $16\%$ and $5\%$, respectively. Since this kind of evolution requires a value of the eccentricity of Ganymede before the resonant encounter slightly larger than today's, it follows that tidal dissipation within the moon must be low. In fact, a null free eccentricity of Ganymede would have led to an almost certain capture into the 7:3 resonance that would have survived until today. As a result of the resonance encounter, the amplitude of the free libration of the Laplace resonant angle increased, providing an explanation for its current value. Moreover, we found evidence that a three-body resonance crossing between the three outer moons occurred within the last few tens of thousands of years.
\end{abstract}

\begin{keywords}
Jupiter, satellites \sep Satellites, dynamics \sep Resonances, orbital \sep Tides, solid body \sep Celestial mechanics
\end{keywords}

\maketitle

\section{Introduction}\label{sec:intr}

The three inner Galilean moons, Io, Europa, and Ganymede, are locked in a three-body mean motion resonance known as Laplace resonance \citep{SINCLAIR_1975,FERRAZ-MELLO_1979,LARI-SAILLENFEST_2024}, whose origin is still uncertain \citep{YODER-PEALE_1981,MALHOTRA_1991,PEALE-LEE_2002,OGIHARA-IDA_2012}. Denoting moons' mean motions with $n_i$ and their mean longitudes with $\lambda_i$ ($i=1,2,3,4$ in order of increasing distance from the planet), the resonant relation reads (on average) $n_1-3n_2+2n_3=0$ and the associated resonant angle $\phi_L=\lambda_1-3\lambda_2+2\lambda_3$ librates around $180^\circ$. Such an orbital configuration forces the eccentricity of the moons to non-zero values, enhancing and maintaining energy dissipation caused by tides on the satellites \citep{PEALE-etal_1979,CASSEN-etal_1979}.

The free libration of the Laplace resonant angle $\phi_L$ was first measured by \citet{LIESKE_1980} through the fit of the so-called Samspon-Lieske analytical theory to astrometric data of the Galilean moons. Later works produced numerical series of their orbital elements directly through $N$-body numerical integrations \citep{MUSOTTO-etal_2002,LAINEY-etal_2006}, confirming and refining the results of Lieske. According to the most recent ephemerides of the Galilean moons (\textit{jup365}\footnote{\url{https://naif.jpl.nasa.gov/pub/naif/generic_kernels/spk/satellites}, last access on 29th June 2026} and \textit{NOE-5}\footnote{\url{https://ftp.imcce.fr/pub/ephem/satel/NOE/JUPITER/2023}, last access on 29th June 2026}), the free libration of the Laplace resonant angle has a proper period of $2060$ days and an amplitude of $0.061^\circ$. As at the time of formation of the resonance $\phi_L$ passed from circulation to libration, its amplitude started from $360^\circ$ and decreased to its current value. Assuming a smooth damping process, previous works used the amplitude of $\phi_L$ to estimate the age of the Laplace resonance in the context of formation driven by tidal migration \citep{YODER-PEALE_1981,HENRARD_1983,MALHOTRA_1991}.

The fourth moon, Callisto, is currently not involved in any mean motion resonance, but it is very close to the 7:3 commensurability with Ganymede, also known as De Haerdtl great inequality \citep{DEHAERDTL_1892}. Despite its high order, such a resonance could have produced significant effects on the orbital elements of the two moons \citep{NOYELLES-VIENNE_2005,NOYELLES-VIENNE_2007}. However, so far, no reconstruction of the evolution of the Galilean moon system through the 7:3 resonance has been conducted, and its past interaction with the Laplace resonance is not known. Since the ratio of the mean motions of Ganymede and Callisto is $n_3/n_4=2.3329$ (mean value), a tiny orbital shift in the semi-major axis of one of the two moons would place them in the 7:3 resonant region (nominally $n_3/n_4=2.3333$).

Tidal forces between Jupiter and its satellites produce secular trends on moons' semi-major axes $a_i$ and eccentricities $e_i$, and drive the evolution after their formation. The magnitude of tidal effects is encoded by the dissipative parameter $k_2/Q$, which is the ratio between the tidal Love number $k_2$ and the quality factor $Q$ of a body (see, e.g., \citealp{KAULA_1964}). Tides on Jupiter make moons' semi-major axes and eccentricities increase, while tides on synchronous satellites make the same orbital elements decrease. As Io is the closest to the planet, tides are prominent on this moon, as evidenced by its strong volcanism \citep{PEALE-etal_1979,VEEDER-etal_1994,LAINEY-etal_2009,DEKLEER-etal_2024}. However, because of the resonant relation, there is a transfer of angular momentum and energy to the other moons involved in the Laplace resonance. As a consequence, Ganymede is the moon with the fastest migration rate, and not Io.

Considering the estimated current migration rate of Ganymede of about $10$ cm/yr \citep{LAINEY-etal_2009} and a negligible orbital expansion of Callisto, it suffices to go back in time for two million years to reach the nominal resonant ratio $n_3/n_4=7/3$. This is an extremely short time interval compared to the billion-year evolution of the system. Therefore, the crossing of the 7:3 mean motion resonance could be the cause, at least in part, of orbital features of the system which are still unexplained, such as the moons' free eccentricities and inclinations, and the libration amplitude of the Laplace resonant angle.

Unlike Io and Europa, the current values of the eccentricities of Ganymede and Callisto show significant free components. More precisely, a bit less than half of the average value of Ganymede's eccentricity (i.e., $0.0007$ out of $0.0016$) is free, while the rest is forced by the resonant and secular perturbations of the other moons. The eccentricity of Callisto is almost entirely free, as the moon is not involved in any mean motion resonance. Moreover, its inclination with respect to its Laplace plane is $0.25^\circ$. As tidal dissipation is expected to damp all these values over a timescale of hundreds of millions of years (see, e.g., \citealp{DOWNEY-etal_2020}), past works tried to explain them through dynamical excitations due to past resonant interactions among the moons \citep{TITTEMORE_1990,MALHOTRA_1991,SHOWMAN-MALHOTRA_1997,DOWNEY-etal_2020,LARI-etal_2023}. However, none of the proposed dynamical pathways is free from inconsistencies when considering the global history of the Galilean moons.

Furthermore, because of the strong tidal dissipation in the Jovian system \citep{LAINEY-etal_2009}, the damping timescale of the amplitude of the free libration of $\phi_L$ is relatively short, and we would expect a present-day value much closer to zero, unless the Laplace resonance formed recently. Looking at the current amplitude, previous works dated the formation of the Laplace resonance around $2\,000\, Q_{\text{J}}$ years ago, where $Q_{\text{J}}$ is the quality factor of Jupiter \citep{YODER-PEALE_1981,HENRARD_1983}. Given the estimated value for $Q_{\text{J}}$ \citep{LAINEY-etal_2009,PARK-etal_2025}, this time would correspond to about $100$ Myr. Although not impossible, such a late formation contradicts the expected tidal heating history of Io \citep{DEKLEER-etal_2024}, and is completely at odds with a primordial origin of the resonance \citep{PEALE-LEE_2002,OGIHARA-IDA_2012}, unless it recently reformed due to some dynamical destabilization (see \citealp{LARI-etal_2023}). Therefore, we investigate whether, instead, the observed libration amplitude is the product of a recent orbital excitation of the system.

In this work, we explore the effects of the 7:3 resonance crossing on the recent orbital history of the Galilean moon system. More precisely, we aim at determining the evolution that best matches the present-day values of the moons' orbital parameters. Moreover, we track the variation of the libration amplitude of the Laplace resonant angle during the resonance crossing.

\section{Methods}\label{sec:meth}

\subsection{Dynamical model}\label{subsec:dynmo}

We considered the gravitational $5$-body model composed by Jupiter and the four Galilean moons: Io (1), Europa (2), Ganymede (3) and Callisto (4). We also included perturbations due to the oblateness of Jupiter ($J_2$ and $J_4$ parameters) and the point-mass perturbation of the Sun. The latter is essential to take into account the non-negligible inclination of the Laplace plane of the outer moons with respect to the equator of Jupiter.

The resulting Hamiltonian, expressed in equatorial jovicentric canonical coordinates (planetocentric positions $\mathbf r_i$ and barycentric momenta $\mathbf p_i=m_i\mathbf v_i$, where $m_i$ and $\mathbf v_i$ denote masses and barycentric velocities, respectively), is \citep{LARI_2018,LARI-etal_2020}
\begin{equation}\label{eq:Htot}
\mathcal{H} = \mathcal{H}_0 + \mathcal{H}_\text{J} + \mathcal{H}_\text{M} + \mathcal{H}_\odot\;,
\end{equation}
where
\begin{equation}\label{eq:Hterms}
   \begin{aligned}
      \mathcal{H}_0 &= \sum_{i=1}^4\left(\frac{\|\mathbf{p}_i\|^2}{2\beta_i} - \frac{\mu_i\beta_i}{\|\mathbf{r}_i\|}\right) \,,\\
      \mathcal{H}_\mathrm{J} &= \sum_{i=1}^4 U_i^{\text{J}}(\mathbf{r}_i) \,,\\
      \mathcal{H}_\mathrm{M} &= - \sum_{1\leqslant i<k\leqslant 4}\left(\frac{\mathcal{G}m_im_k}{\|\mathbf{r}_i-\mathbf{r}_k\|} - \frac{\mathbf{p}_i\cdot\mathbf{p}_k}{m_\text{J}}\right) \,,\\
      \mathcal{H}_\odot &= - \sum_{i=1}^4\frac{\mathcal{G}m_im_\odot}{\|\mathbf{r}_i-\mathbf{r}_\odot\|} + \ddot{\mathbf{x}}_\text{G}\cdot\sum_{i=1}^4 m_i\mathbf{r}_i \,,\\
   \end{aligned}
\end{equation}
which correspond to the Keplerian, oblateness, mutual perturbation, and solar terms, respectively. In Eq.~\eqref{eq:Hterms}, $\beta_i=m_\text{J}m_i/(m_\text{J}+m_i)$ are the reduced masses ($m_\text{J}$ the mass of Jupiter); $\mu_i=\mathcal G(m_\text{J}+m_i)$ are the gravitational parameters ($\mathcal G$ the gravitational constant); $U_i^{\text{J}}$ is the gravitational potential energy associated with the second ($J_2$) and fourth ($J_4$) zonal harmonic coefficients of Jupiter (see, e.g., \citealp{KAULA_1966} for its expression); and $\ddot{\mathbf{x}}_\text{G}$ is the gravitational acceleration of Jupiter in an inertial frame due to the Sun (mass $m_{\odot}$ and position $\mathbf{r}_\odot$). In our model, we considered a fixed orbit for Jupiter set to its values at J2000 epoch. The equations of motion were obtained from the classic Hamilton's equations.

\begin{table}
\caption{Physical parameters of Jupiter and the Galilean moons.}\label{tbl1}
\begin{tabular*}{\tblwidth}{@{}LCR@{}}
\toprule
parameter  & unit  &  value \\ 
\midrule
$\mathcal{G}$ & $R_\text{J}^3/(m_\text{J}\text{yr}^2)$ & $3.45275721\times 10^{8}$\\
$J_2$ & & $1.46965063\times 10^{-2}$\\
$J_4$ & & $-5.866085\times 10^{-4}$ \\
$m_1$ & $m_\text{J}$ & $4.70445861\times 10^{-5}$\\
$m_2$ & $m_\text{J}$ & $2.52806044\times 10^{-5}$\\
$m_3$ & $m_\text{J}$ & $7.80495969\times 10^{-5}$\\
$m_4$ & $m_\text{J}$ & $5.66696656\times 10^{-5}$\\
$m_\odot$ & $m_\text{J}$ & $1.04757176\times 10^{3}$\\
$R_1$ & $R_\text{J}$ & $2.55888770\times 10^{-2}$\\
$R_2$ & $R_\text{J}$ & $2.18569909\times 10^{-2}$\\
$R_3$ & $R_\text{J}$ & $3.68041179\times 10^{-2}$\\
$R_4$ & $R_\text{J}$ & $3.37142617\times 10^{-2}$\\
\bottomrule
\end{tabular*}

\vspace{1mm}
\begin{minipage}{\linewidth}
Values are taken from the spice kernel \textit{jup365}. Reference values for $R_\text{J}$, $m_\text{J}$, and yr are $71\,492$ km, $1.89812462\times 10^{27}$ kg and $31\,557\,600$ s, respectively.
\end{minipage}
\end{table}

We added tidal forces directly to the differential equations following the formulation of \cite{MIGNARD_1979}. The force acting on the $i$-th moon as a result of the tides on Jupiter can be expressed as
\begin{multline}
   \label{eq:tidemignard}
   \mathbf F_{T,\text{J}}=-3\left(\frac{k_{2}}{Q}\right)_\text{J}\frac{1}{2(\omega_\text{J}-n_i)}\frac{\mathcal{G}m_i^2R_\text{J}^5}{r_i^7}\times \\
   \left(2\frac{\mathbf{r}_i\cdot\mathbf{v}_i}{r_i^2}\frac{\mathbf{r}_i}{r_i}+\frac{\mathbf r_i \times \boldsymbol \omega_\text{J} + \mathbf v_i}{r_i}\right)\;,
\end{multline}
where $\boldsymbol \omega_\text{J}$, $R_\text{J}$ are the spin vector and the radius of Jupiter, respectively, and $r_i=\|\mathbf{r}_i\|$. In order to include the effects of moon tides, we also considered the following forces, which are derived imposing synchronous rotation of the satellites \citep{LARI_2018}:
\begin{equation}
   \label{eq:tidelari}
   \mathbf F_{T,i}=-21\left(\frac{k_2}{Q}\right)_i\frac{1}{n_i}\frac{\mathcal{G}m_\text{J}^2R_i^5}{r_i^7}\frac{\mathbf{r}_i}{r_i}
   \frac{\mathbf{r}_i\cdot\mathbf{v}_i}{r_i^2}\;,
\end{equation}
where $R_i$ are the radii of the satellites $(i=1,2,3,4)$. All values of the physical parameters that we used are reported in \tablename~\ref{tbl1}.

Eqs.~\eqref{eq:tidemignard} and \eqref{eq:tidelari} reproduce qualitatively and quantitatively the secular behavior described by the classical formulas for semi-major axes $a_i$ and eccentricities $e_i$ \citep{KAULA_1964,PEALE-etal_1979}:
\begin{equation}
\label{eq:tides}
\begin{aligned}
\frac{\dot{a}_i}{a_i}&=\frac23 
c_i\left(1 -7D_ie_i^2\right)\;,\\
\frac{\dot{e}_i}{e_i}&=-\frac13c_i\left(7D_i-\frac{19}{4}\right)\;,
\end{aligned}
\end{equation}
with
\begin{equation}
\label{eq:ciDi}
\begin{aligned}
c_i&=\frac92 \left(\frac{k_{2}}{Q}\right)_\text{J}\frac{m_i}{m_\text{J}}\left(\frac{R_\text{J}}{a_i}\right)^5n_i\;, \\
D_i&=\left(\frac{k_{2}}{Q}\right)_i\left(\frac{Q}{k_2}\right)_\text{J}\left(\frac{R_i}{R_\text{J}}\right)^5\left(\frac{m_\text{J}}{m_i}\right)^2\;.
\end{aligned}
\end{equation}

The tides raised by a planet on a synchronous satellite moving on eccentric orbit generate friction which dissipate energy at the following rate \citep{PEALE-CASSEN_1978,PEALE-etal_1979}:
\begin{equation}
   \label{eq:Ediss}
   \frac{dE}{dt}=-\frac{21}{2}\left(\frac{k_{2}}{Q}\right)_i\frac{n_i^5R_i^5}{\mathcal G}e_i^2\;.
\end{equation}
From this formula, we see how energy dissipation depends on the eccentricity and semi-major axis of the satellite. For Io, we compute a tidal heating of approximately $10^{14}$ W \citep{LAINEY-etal_2009}. During the billion-year history of the Galilean moons, their tidal heating could have changed significantly because of the variation of their orbital elements. In particular, if the moons were temporarily captured into resonances different from the Laplace resonance, their eccentricities could have increased enough to induce geophysical effects like tectonic activity and resurfacing also on the outer moons (see, e.g., \citealp{MALHOTRA_1991,SHOWMAN-MALHOTRA_1997}).

\subsection{Tidal parameters}
We considered constant dissipative parameters $k_2/Q$ for all the involved bodies. This is a reasonable assumption considering the relatively short time interval of the evolution (a few million years). For Jupiter, we set a value of $1.3\times 10^{-5}$, equal at all moons' orbital frequencies, which is compatible with the estimation obtained from astrometric observations \citep{LAINEY-etal_2009}.

We then chose a dissipative parameter of Io in such a way that the equilibrium eccentricities of Io and Europa are equal to their current mean values ($0.0042$ and $0.0094$, respectively). An analytical approximation of the equilibrium eccentricity of Io as a function of the tidal parameters is given by \citet{YODER-PEALE_1981}. Their formula yields $k_2/Q=0.011$, which is exactly the value estimated from Juno data \citep{PARK-etal_2025} and is also not far from the previous astrometric estimate \citep{LAINEY-etal_2009}. This choice makes the comparison of our results with the current configuration of the system straightforward.

For the other moons, no direct estimations are available; therefore, some assumptions were made. For Europa, we considered a $k_2/Q$ of $0.01$, which reflects its expected significant tidal heating. For Ganymede, we could consider either a high or low $k_2/Q$, but the choice must be consistent with the initial value of its eccentricity. If we set a dissipative parameter similar to those of Io and Europa, then the timescale of eccentricity damping would be around $25$ million years, implying that Ganymede's free eccentricity before the resonance crossing would have been completely damped. Conversely, if we assume a low $k_2/Q$ for Ganymede, then the pre-resonance value of its eccentricity could have been similar or even slightly higher than today's. In our setup, we assumed the latter, setting $k_2/Q=0.001$. However, in Sect.~\ref{sec:disc}, we discuss also the case in which tidal dissipation within Ganymede is high.

For Callisto, the choice is quite arbitrary. In fact, because of its distance from Jupiter, tides on Callisto are expected to be extremely small and they produce significant effects on its orbital elements only over hundreds of millions of years at best, which is about two orders of magnitude larger than the timescale considered in this study. Therefore, we set its $k_2/Q$ equal to $0.001$, similarly to Ganymede.

\subsection{Initial conditions}

We considered a pre-resonance orbital configuration almost identical to the current one, apart from the eccentricities of Ganymede and Callisto, and the inclination of Callisto. We took values of Ganymede's eccentricity slightly larger than its current value, i.e., between $0.00175$ and $0.00205$ (average value) to account for downward kicks at the resonance crossing. For Callisto, we considered both the case of a slightly larger eccentricity, i.e., between $0.0077$ and $0.0080$, and a smaller eccentricity, i.e., $<0.0070$, to explore evolution without and with capture into resonance, respectively. For the same reason, we took the inclination of Callisto equal to or smaller than its current value, i.e., between $0.05^\circ$ to $0.25^\circ$ with respect to its Laplace plane.
 
Moreover, considering the migration rates of the moons, we placed the three inner moons slightly interior to their current orbits, to start the simulations around two million years before the 7:3 resonance crossing (i.e., $n_3/n_4\approx 2.334$, mean value). This choice of initial $a_3/a_4$ places the current epoch approximately four million years after the start of the numerical integration, if the moons were not captured into resonance.

In order to start the numerical simulations with a near-zero amplitude of the free libration of the Laplace resonant angle, we ran preliminary simulations of the system using a large acceleration factor on tidal effects \citep{MALHOTRA_1991,LARI-etal_2020}, which quickly damped the amplitude of $\phi_L$. Initial conditions were then taken from the output of these simulations (see the example reported in \tablename~\ref{tbl2}).

Apart from varying the initial value of the eccentricity of Ganymede and Callisto, and the inclination of Callisto, in order to take into account the stochasticity of the resonance crossing, we ran 25 different simulations for each different setup (i.e., for chosen values of $e_3$, $e_4$, and $I_4$), taking slightly different values of the longitude of Callisto $\lambda_4$. We changed this orbital element by a small amount ($0.0001$ radians) among simulations to avoid exciting $\phi_L$. Over one million years of propagation, this small difference in the initial condition spreads across $2\pi$ radians, generating different angular phases at the resonance encounter.

\begin{table}
\caption{Initial orbital conditions for one of our numerical simulations.}\label{tbl2}
\begin{tabular*}{\tblwidth}{@{}LCR@{}}
\toprule
parameter & unit & value  \\ 
\midrule
$a_1$ & $R_\text{J}$ & $5.90009200$\\
$a_2$ & $R_\text{J}$ & $9.38493859$\\
$a_3$ & $R_\text{J}$ & $14.96984483$\\
$a_4$ & $R_\text{J}$ & $26.33513796$\\
$e_1$ & & $0.00432182$\\
$e_2$ & & $0.00910933$\\
$e_3$ & & $0.00136230$\\
$e_4$ & & $0.00789036$\\
$I_1$ & rad & $0.00076015$\\
$I_2$ & rad & $0.00822694$\\
$I_3$ & rad & $0.00187371$\\
$I_4$ & rad & $0.00950688$\\
$\omega_1$ & rad & $2.12077359$\\
$\omega_2$ & rad & $4.47402221$\\
$\omega_3$ & rad & $2.11556817$\\
$\omega_4$ & rad & $0.57424328$\\
$\Omega_1$ & rad & $1.58745487$\\
$\Omega_2$ & rad & $2.28941490$\\
$\Omega_3$ & rad & $1.42926015$\\
$\Omega_4$ & rad & $5.73594574$\\
$\ell_1$ & rad & $1.13705272$\\
$\ell_2$ & rad & $3.75319664$\\
$\ell_3$ & rad & $1.95344481$\\
$\ell_4$ & rad & $3.03224549$\\
\bottomrule
\end{tabular*}

\vspace{1mm}
\begin{minipage}{\linewidth}
The (osculating) orbital elements are: semi-major axis ($a$), eccentricity ($e$), inclination ($I$), argument of the pericenter ($\omega$), longitude of the ascending node ($\Omega$), and mean anomaly ($\ell$).
\end{minipage}
\end{table}

\subsection{Numerical integration}
Unlike previous works \citep{MALHOTRA_1991,SHOWMAN-MALHOTRA_1997,LARI-etal_2020,LARI-etal_2023}, we did not apply any acceleration factor on tidal effects to speed up the evolution during the resonance crossing. In fact, in contrast to the 2:1 resonant chains explored in those articles, the 7:3 mean motion resonance is relatively weak. Increasing the magnitude of tides would not simply rescale the evolution time, but it would also introduce artificial effects in the dynamics.

For our investigation, we used the open-source N-body code \texttt{REBOUND} \citep{REIN-LIU_2012}. In addition, we incorporated the \texttt{REBOUNDx} library to account for additional physical effects \citep{TAMAYO-etal_2020}, including Jupiter’s oblateness, solar perturbation, and tidal dissipation. To adapt the code to our model, we added tidal forces to the \texttt{REBOUNDx} library according to Eqs. \eqref{eq:tidemignard} and \eqref{eq:tidelari}. The implemented setup is equivalent to the dynamical model presented in Sect.~\ref{subsec:dynmo}, which can also be propagated outside \texttt{REBOUND}.

Simulations using \texttt{REBOUND} were performed with the \texttt{WHFAST} integrator \citep{REIN-TAMAYO_2015}, based on an implementation of the Wisdom–Holman symplectic method \citep{WISDOM-HOLMAN_1991} for long-term orbital integrations. The additional tidal forces were implemented as \texttt{REBOUNDx} operators, which allow the particle velocities to be updated without interfering with the symplectic nature of the \texttt{WHFAST} integrator. Because the timescale of tidal dissipation is much longer than the secular and resonant timescales, the dynamics can be regarded as predominantly conservative, with small dissipative perturbations that do not significantly alter the overall behavior of the numerical scheme.

To verify this, we compared the results obtained with \texttt{WHFAST} with those produced by the high-accuracy \texttt{IAS15} integrator \citep{EVERHART_1985,REIN-SPIEGEL_2015}. For \texttt{WHFAST}, we adopted a step of $2.2\times 10^{-4}$ yr, which was found to provide a good compromise between accuracy (comparable to \texttt{IAS15}) and computational efficiency (5–6 times faster).

\section{Results}\label{sec:resu}
We ran hundreds of numerical simulations of the resonance crossing with the setup described in Sect.~\ref{sec:meth}. The simulations yielded different outcomes: in some cases, Ganymede and Callisto were not captured into the 7:3 resonance, whereas in others they were. In this section, we show how the first scenario manages to perfectly match the current orbital configuration of the system. For the second scenario, temporary capture into resonance could be possible in principle. However, in most simulations of this kind, the two outer moons were captured into some 7:3 sub-resonances that persisted enough to increase their eccentricities well above their present-day values, making such an evolution not compatible with the current configuration of the system. We discuss this last case in Sect.~\ref{sec:disc}.

\subsection{Evolution without capture}\label{sec:res1}

\begin{figure}
\centering
\includegraphics[scale=0.6]{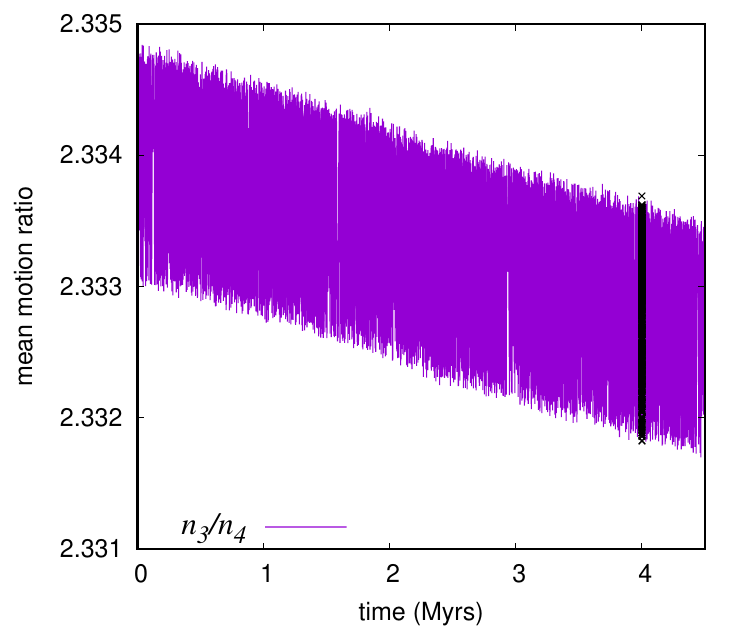}
\caption{Evolution of the mean motion ratio between Ganymede and Callisto in the case they were not captured into the 7:3 resonance. In black, the evolution of the same quantity obtained starting the propagation from initial conditions taken from ephemerides at J2000 epoch.}
\label{fig:a34}
\end{figure}

We ran in total $250$ numerical simulations starting from eccentricities of Ganymede and Callisto slightly larger than their present-day values (see Sect.~\ref{sec:meth}). We found that in $65\%$ of the simulations, there was no capture into the 7:3 resonance. The outcomes of this type of simulations are very similar and align with the current orbital configuration of the system, as we show below.

Given the initial conditions and tidal parameters as described in Sect.~\ref{sec:meth}, the mean motion ratio $n_3/n_4$ decreases at a rate of $-3\times 10^{-10}$ per year. Therefore, the crossing of the 7:3 resonance occurred around two million years from the beginning of our simulations. Since in these simulations the moons were not captured into resonance, the mean motion ratio shows a constant trend for the whole evolution, and after about four million years the system reached the current orbital proportion (i.e., $n_3/n_4=2.3329$, mean value), which sets the present-day epoch (see \figurename~\ref{fig:a34}).

\begin{figure*}
\centering
\includegraphics[scale=0.6]{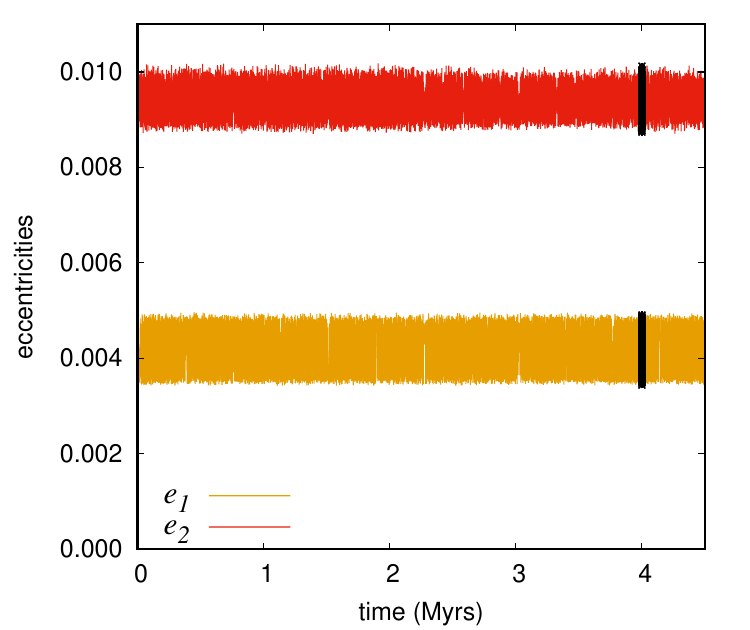}\includegraphics[scale=0.6]{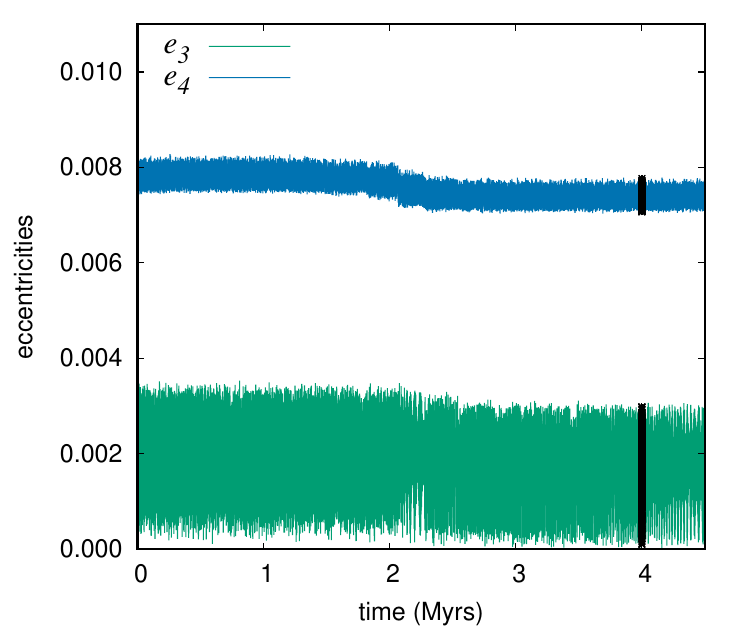}
\caption{Evolution of the eccentricities of Io and Europa (left), and of Ganymede and Callisto (right) in the case the latter were not captured into the 7:3 resonance. In black, the evolution of the same elements obtained starting the propagation from initial conditions taken from ephemerides at J2000 epoch.}
\label{fig:ecc}
\end{figure*}

\begin{figure}
\centering
\includegraphics[scale=0.6]{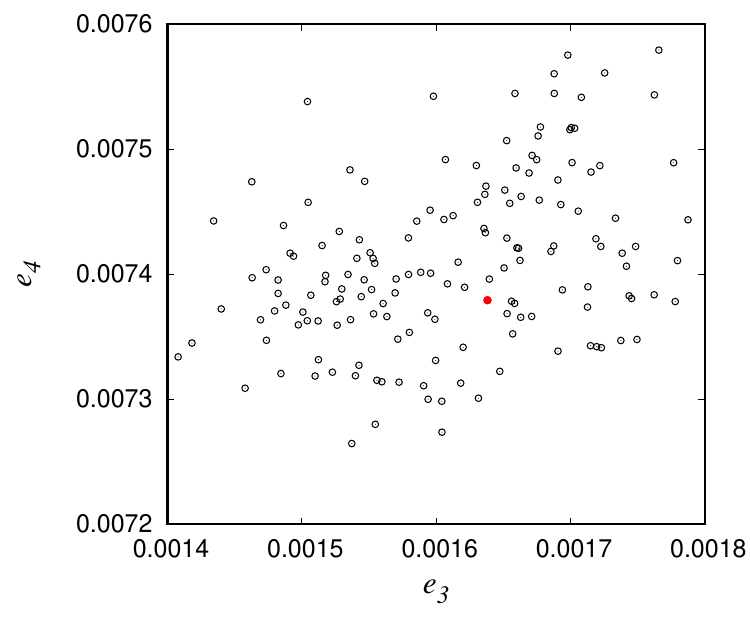}
\caption{Final values of the mean eccentricities of Ganymede and Callisto for all simulations without capture into resonance. The red dot represents the simulation reported in \figurename~\ref{fig:ecc}.}
\label{fig:kicke34}
\end{figure}

The crossing of the resonance without capture produced a downward kick in the eccentricities of both Ganymede and Callisto (see \figurename~\ref{fig:ecc}, right panel). This behavior is common in convergent evolution when the system passes through the separatrices of one or more sub-resonances (e.g., \citealp{MURRAY-DERMOTT_2000}). The average eccentricity of Ganymede decreased by $16\%\pm 2\%$ and the average eccentricity of Callisto by $5\%\pm 0.5\%$. The reported standard deviation describes the different kick sizes between simulations as a result of the stochasticity of the resonance crossing. Through the initial tuning of the pre-resonance values of the eccentricities, we managed to retrieve final values of these orbital elements close to the current ones. More precisely, for all simulations without capture, we obtained a mean post-resonance value (with standard deviation) of $e_3$ of $0.0016\pm 0.0001$, and a mean post-resonance value of $e_4$ of $0.0074\pm 0.0001$ (see \figurename~\ref{fig:kicke34}). As we assumed small dissipative parameters $k_2/Q$ of Ganymede and Callisto, the tidal damping in their eccentricities was almost negligible throughout the evolution.

Unlike Ganymede and Callisto, the eccentricities of Io and Europa remained almost unaffected by the resonant encounter (see \figurename~\ref{fig:ecc}, left panel). In fact, the 7:3 resonance caused only an indirect perturbation on the two inner moons through the (small) change of the eccentricity of Ganymede. Therefore, at the end of the simulations, the moons always reached the correct values for both $e_1$ and $e_2$.

Finally, all inclinations remained almost unchanged through the resonance crossing. The resulting system then closely matches the current orbital configuration of the Galilean moons. As this kind of evolution was obtained in about two thirds of the simulations, this is the most likely pathway followed by the moons through the 7:3 resonance.

\subsection{Temporary capture into resonance}\label{sec:res2}

It is also possible for Ganymede and Callisto to be captured into the 7:3 resonance. Such a resonance comprehends a large number of sub-resonances \citep{NOYELLES-VIENNE_2005}, so that the evolution of the moons' orbital elements depends strongly on which one is triggered by the moons.

The first to be encountered are sub-resonances in inclinations, the latest in eccentricities (see, e.g., \citealp{MURRAY-DERMOTT_2000}), while the mixed-type ones are spread over the whole resonant region. The moons can skip certain sub-resonances and be trapped into others. The capture probability depends on resonance efficiency, moons' migration rate, initial orbital eccentricities and inclinations, and a certain degree of stochasticity due to the different angular phases with which the resonances are encountered.

A detailed study of all possible 7:3 sub-resonances is out of the scope of this work, so that in this section we focus only on sub-resonances whose effects are in principle compatible with the current configuration of the system. This is not the case of sub-resonances involving the eccentricity of Ganymede, like $e_3e_4^3$ and $e_3^2e_4^2$, as they cause a rapid increase in its value. Since after the eventual disruption of the 7:3 resonance there is not enough time to damp $e_3$ and $e_4$, we discarded such sub-resonances from the possible pathways followed by the system. More details are given in Sect.~\ref{sec:disc}.

Instead, temporary capture into sub-resonances affecting only the eccentricity and inclination of Callisto could allow to match the current configuration of the system. Both elements show a significant free component (see Sect.~\ref{sec:intr}), so that it is possible that Callisto obtained such values, or at least part of them, during the passage through the 7:3 resonance. We then ran dedicated simulations starting from values of $e_4$ and $I_4$ smaller than today's, i.e., between $0.004$ and $0.007$, and between $0.05^\circ$ and $0.25^\circ$, respectively.

For most simulations for which a capture occurred, the resonance persisted for many millions of years, causing the eccentricity of Callisto to increase to values larger than $0.008$, above its current average value of $0.0074$. However, in a few simulations, the 7:3 resonance broke down prematurely, resulting in a final value of $e_4$ compatible with the current orbit of Callisto.

\begin{figure*}
\centering
\includegraphics[scale=0.6]{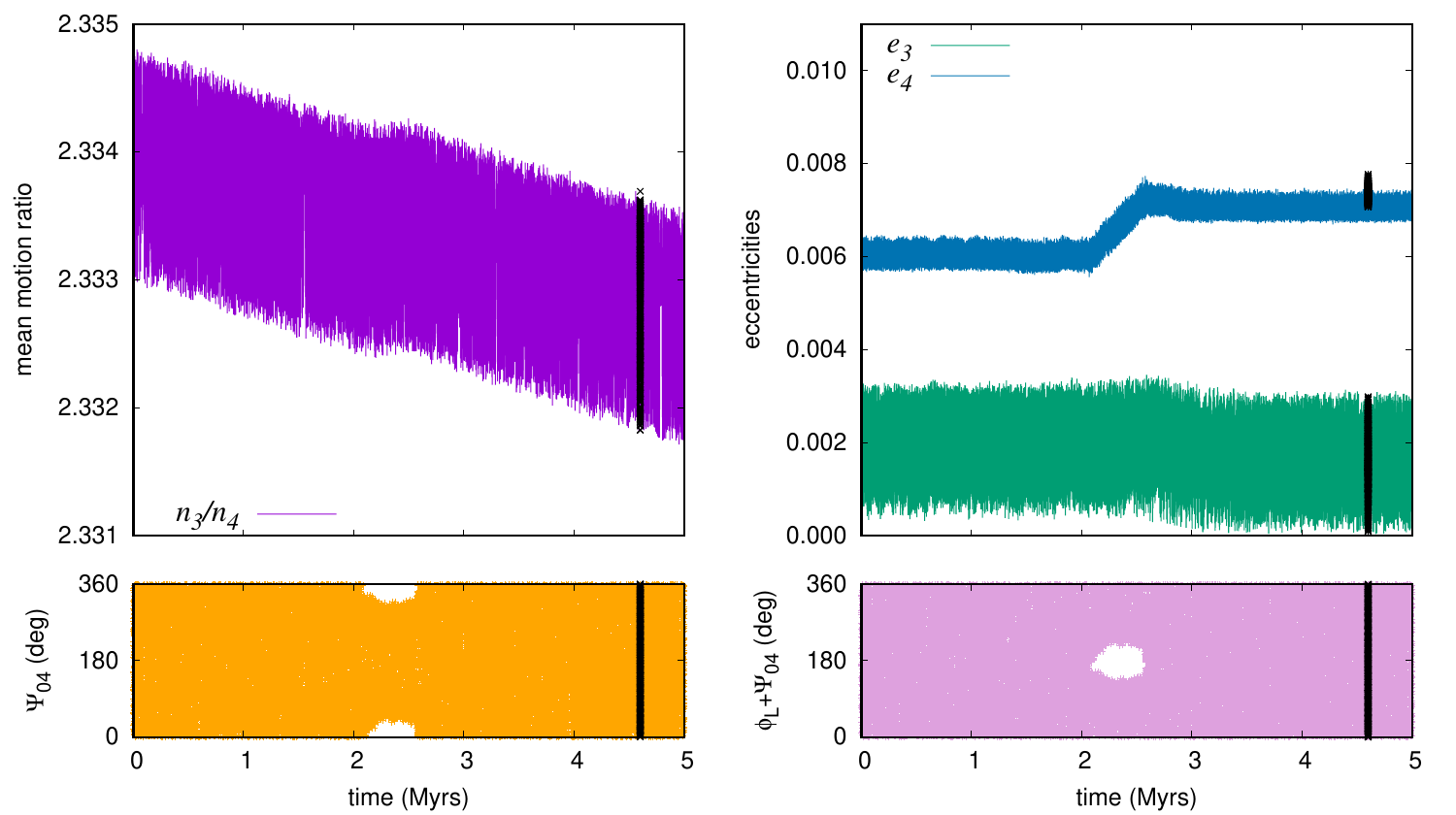}
\caption{Evolution of the system in the case of a temporary capture into the $e_4^4$ sub-resonance: mean motion ratio between Ganymede and Callisto (top left), eccentricities of Ganymede and Callisto (top right), resonant argument $\Psi_{04}$ (bottom left), and four-body geometric angle $\phi_L+\Psi_{04}$ (bottom right). In black, the evolution of the same quantities obtained starting the propagation from initial conditions taken from ephemerides at J2000.}
\label{fig:temcap}
\end{figure*}

In \figurename~\ref{fig:temcap}, we reported a case where the 7:3 resonance lasted about $0.5$ Myrs, and the eccentricity of Callisto increased from a mean value of $0.0061$ to about $0.0071$, i.e., close to its present value. In this particular simulation, the moons triggered the $e_4^4$ sub-resonance, whose corresponding resonant argument is $\Psi_{04}=3\lambda_3-7\lambda_4+4\varpi_4$. In some other simulations, we obtained an increase in both the eccentricity and inclination of Callisto, but they were rarer and often involved also $I_3$. Once captured into resonance, also Callisto was pushed outwards by the tidal effects between Jupiter and Io, and the ratio $n_3/n_4$ stabilized on a constant value close to $7/3$.

An early exit from the resonance can depend on many factors, for example the proximity to the separatrix associated with the resonance after capture. In general, most simulations where the resonance broke down at $e_4\le 0.008$ started already with a high initial value of $e_4$ (like in \figurename~\ref{fig:temcap}), and the increase of the eccentricity of Callisto is limited to a few units of $0.0001$. Therefore, a temporary capture into such sub-resonance could have occurred, but its overall impact in the orbital evolution of the system would have been quite limited. Moreover, the probability of capture into sub-resonances not involving $e_3$ is very small, about $10-15\%$ for initial $e_4\ge0.006$, and an early disruption is quite rare.

In the end, finding a simulation that perfectly matches the current value of $e_4$ requires both careful tuning of the initial conditions and a certain amount of luck to get the exact duration of the resonance. It follows that this kind of evolution is far less probable than the case of crossing without capture described in Sect.~\ref{sec:res1}, and that it is very likely that Ganymede and Callisto were not captured into the 7:3 resonance, not even temporarily.

\subsection{Laplace resonant angle}

\begin{figure*}
\centering
\includegraphics[scale=0.6]{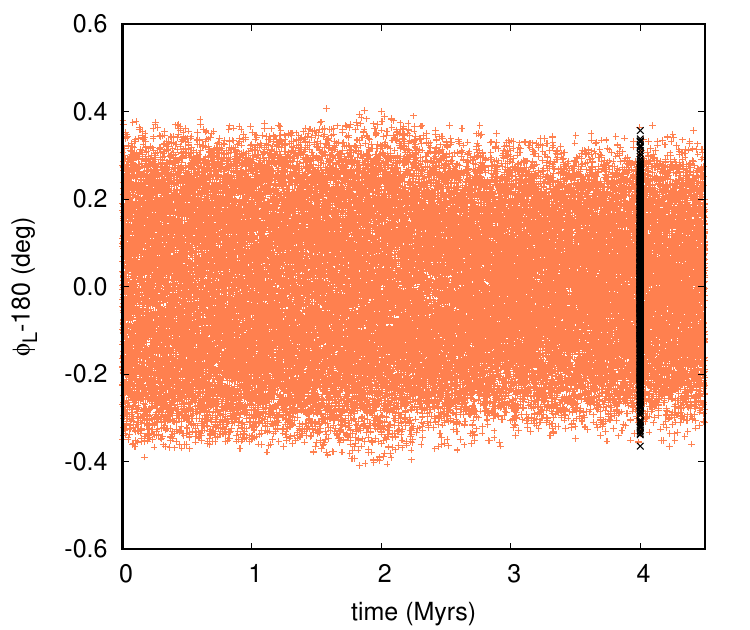}\includegraphics[scale=0.6]{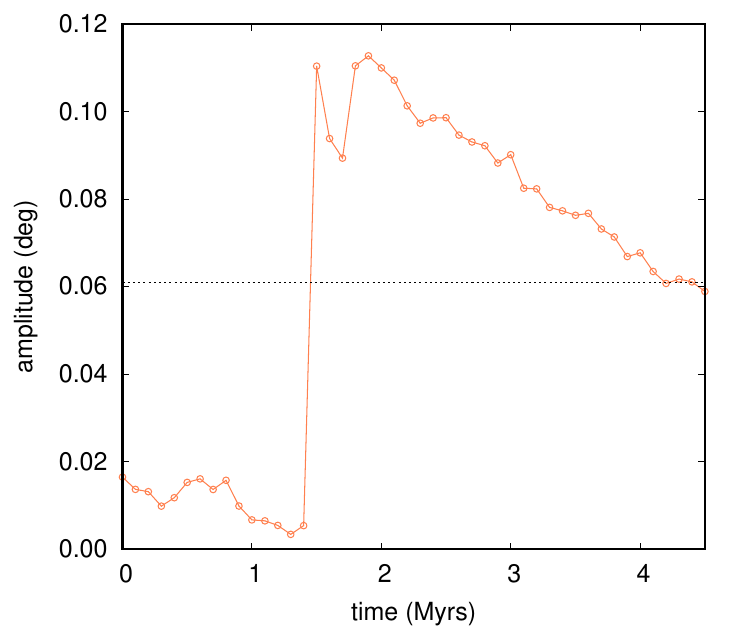}
\caption{Evolution of the Laplace resonant angle (left), and of the amplitude of its free libration (right). In black, the evolution of the same angle obtained starting the propagation from initial conditions taken from ephemerides at J2000 epoch. The dashed line on the right corresponds to the current value of the amplitude.}
\label{fig:laplace}
\end{figure*}

One effect of the recent crossing of the 7:3 resonance was to excite the Laplace angle $\phi_L$ associated with the three-body mean motion resonance among Io, Europa, and Ganymede. While the overall excitation appears quite limited (see \figurename~\ref{fig:laplace}, left panel), the amplitude of the free libration shows a remarkable increase (same figure, right panel). Since in the time series of $\phi_L$ the free libration is concealed by other frequencies, we had to extract its signal through a frequency analysis. For this task, we used the program \texttt{giffv} provided in the OrbFit software\footnote{\url{https://adams.dm.unipi.it/orbfit}, last access on 29th June 2026}. We then computed the value of the libration amplitude over a period of $200$ years every $100\,000$ years, tracking its change in value from the start of the evolution to its end.

As described in Sect.~\ref{sec:meth}, we started with initial conditions of the moons' orbits such that the amplitude of the free libration of $\phi_L$ was much smaller than its current value of $0.061^\circ$. As shown in \figurename~\ref{fig:laplace}, the crossing of the 7:3 resonance induced a jump in amplitude, generally exceeding the present-day value. After this upward kick, the amplitude was damped by tidal dissipation within the system.

In \figurename~\ref{fig:laplace}, we reported the evolution of the libration amplitude of $\phi_L$ for the same numerical simulation presented in \figurename~\ref{fig:a34} and~\ref{fig:ecc}. For this simulation, we see that the amplitude at the current epoch (reached after 4 Myr from the beginning of the simulation) is almost exactly the one measured today.

\begin{figure}
\centering
\includegraphics[scale=0.6]{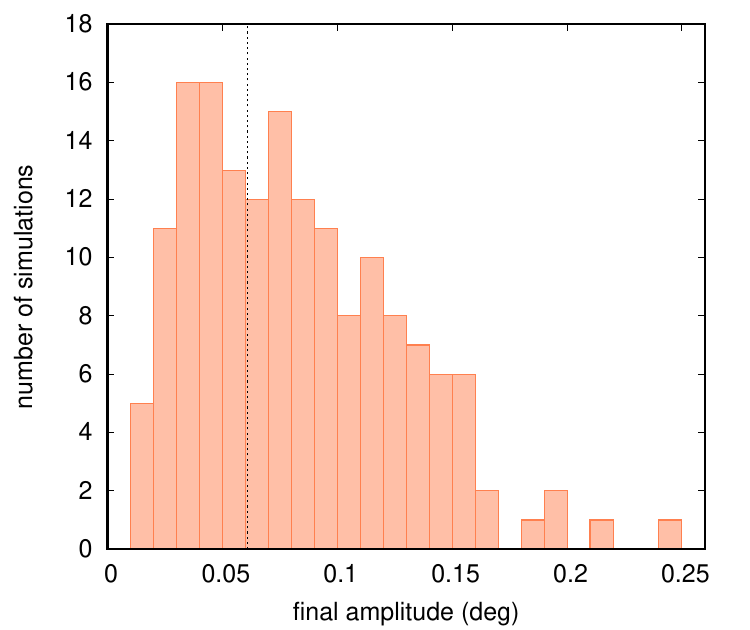}
\caption{Statistics of the final libration amplitude of the Laplace resonant angle obtained from all simulations where Ganymede and Callisto were not captured into resonance. The dashed black line indicates the present-day value.}
\label{fig:hist}
\end{figure}

More generally, the increase in amplitude due to the 7:3 resonance crossing is a stochastic phenomenon. In \figurename~\ref{fig:hist}, we reported the free libration amplitude at the current epoch obtained from all simulations where Ganymede and Callisto were not captured into the 7:3 resonance, not even temporarily. We found that, in most simulations, the final amplitude ranged between $0.02^\circ$ and $0.15^\circ$, with a significant number of simulations yielding values close to the present value of $0.061^\circ$.

\subsection{A last three-body resonance kick}

In our simulations, we observed a final small dynamical excitation of the system around 20 thousands of years before the J2000 epoch. Such a timing is purely indicative, as it depends on the exact values of the dissipative parameters and on the exact identification of the J2000 epoch through the value of $n_3/n_4$. Nevertheless, this event is interesting both because it is extremely recent and its source was not immediately clear.

\begin{figure*}
\centering
\includegraphics[scale=0.6]{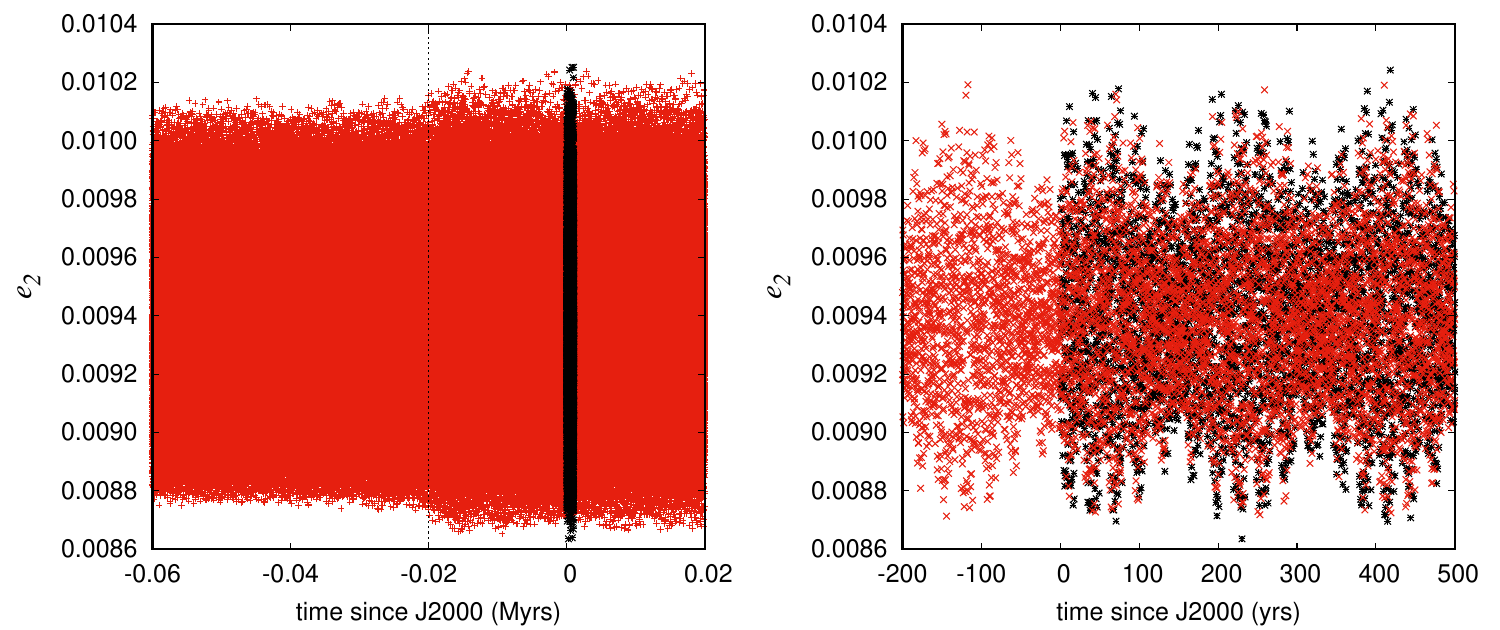}
\caption{Evolution of the eccentricity of Europa during the last tens of thousands of years (left) and during the last few hundred years (right). In black, the evolution of the same element obtained starting the propagation from initial conditions taken from ephemerides at J2000 epoch. The dashed line on the left indicates the time of the last orbital excitation.}
\label{fig:last}    
\end{figure*}

The dynamical excitation is most evident in the eccentricity of Europa, and, as a result, the final oscillation closely matches the current evolution of $e_2$ (see \figurename~\ref{fig:last}). From this observation, we investigated whether this event could have been caused by the crossing of a resonance involving this moon.

The best candidates are three-body resonances among Europa, Ganymede, and Callisto. As shown by \citet{LARI-etal_2020}, such resonances surround a wide region around the nominal two-body resonance. However, it is not straightforward to identify the exact resonance involved. In our exploration, we used the fundamental frequencies ($n_i,g_i,s_i$ for $i=1,2,3,4$), of the system presented in \citet{LAINEY-etal_2006} to search for this new resonant combination. Such values were computed at J2000 epoch, but they can also be considered accurate in an interval of tens of thousands of years. For example, for the 7:3 resonance that the system crossed around two million years ago, the combination $3n_3-7n_4\approx -0.285$ rad/yr. When looking at possible three-body resonant combinations, we found that nowadays the combination $3n_3-7n_4+g_2+3g_4\approx -0.002$ rad/yr. Therefore, such a three-body resonance involving Europa, Ganymede, and Callisto is much closer than the nominal location of the 7:3 mean motion resonance. This three-body resonance is of second order in the masses and of sixth order in the eccentricities. It can be obtained as a combination of the resonant term $3\lambda_3-7\lambda_4+\varpi_3+3\varpi_4$ of the mutual gravitational expansion between Ganymede and Callisto, and the secular term $\varpi_2-\varpi_3$ of the mutual gravitational expansion between Europa and Ganymede.

It is worth noting that $g_2$ is not equivalent to $\dot \varpi_2$. In fact, the evolution of $\varpi_2$ is forced by the Laplace resonance, so that its main frequency is not $g_2$, but $n_1-2n_2$ (see, e.g., \citealp{LAINEY-etal_2006,LARI_2018}). In contrast, since Callisto is not involved in any resonance, the main frequency of $\varpi_4$ is the fundamental frequency $g_4$, so that $\dot \varpi_4 \approx g_4$.

\begin{figure}
\centering
\includegraphics[scale=0.6]{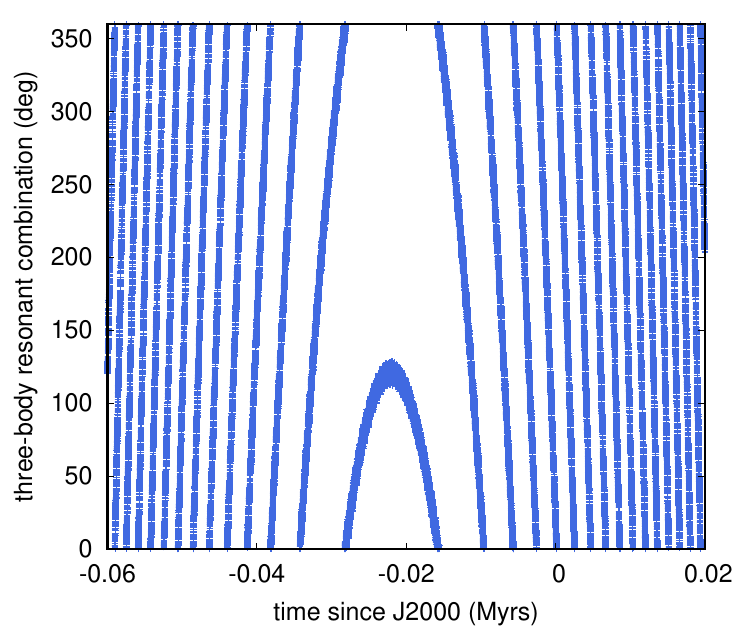}
\caption{Evolution of the three-body resonant argument $3\lambda_3-7\lambda_4+g_2t+3\varpi_4$ during the last tens of thousands of years. The change in sign of the circulation rate marks the crossing of the resonance.}
\label{fig:3b}    
\end{figure}

When plotting the corresponding three-body resonant argument, which is $3\lambda_3-7\lambda_4+g_2t+3\varpi_4$ (see \figurename~\ref{fig:3b}), we can appreciate how around $20\,000$ years ago the angle rate changed sign. This indicates that the system passed through that precise resonance, which exposes its role in the last excitation observed in \figurename~\ref{fig:last}.

It is interesting to note that if the eccentricity of Callisto had significantly increased because of a previous capture into the 7:3 resonance, then it would have been quite probable for the moons to be trapped into this three-body resonance. Even though such a capture could not have occurred in the actual evolution of the Galilean moon system, this highlights the complex dynamics in resonant chains of celestial bodies and the difficulties in foreseeing their dynamical pathways.

\section{Discussion}\label{sec:disc}

In this section, we discuss the case in which the dissipative parameter of Ganymede is higher than the one assumed in our study. Moreover, we present simulations where Ganymede and Callisto are captured into the 7:3 resonance, and we show why this kind of evolution is not compatible with the current configuration of the system.

\subsection{Tidal dissipation within Ganymede}\label{sec:dis1}

The current tidal heating within Ganymede depends on the value of its dissipative parameter and is described by Eq.~\eqref{eq:Ediss}. In our nominal setup, we considered a low tidal dissipation within Ganymede ($k_2/Q=0.001$), so that we could assume that the free eccentricity of Ganymede before the resonant encounter was similar to its current value. The timescale of damping in eccentricity for a synchronous moon is given by
\begin{equation}
\label{eqn:taue}
\tau_e=\frac{2}{21}\left(\frac{Q}{k_2}\right)_i\frac{m_i}{m_\text{J}}\left(\frac{a_i}{R_i}\right)^5 \frac{1}{n_i}
\end{equation}
which corresponds to the interval for the eccentricity to decrease by a factor $\exp(1)$. In Eq.~\eqref{eqn:taue}, the dissipative parameter is the one of the satellite, so that for Ganymede's $k_2/Q=0.001$, we obtain $\tau_e \approx 250$ Myrs.

In Sect.~\ref{sec:res1}, we showed that the most probable evolution resulted to be a crossing of the 7:3 resonance without capture, which produced a downward kick of the eccentricities of Ganymede and Callisto.

However, one could assume that the dissipative parameter of Ganymede is actually high (i.e., $k_2/Q\gtrsim 0.01$). If this is the case, then the damping timescale of the eccentricity would be a few tens of millions of years. Therefore, the free eccentricity of Ganymede before the 7:3 resonance crossing would have been almost null, and the moon should have acquired it through a temporary capture into the 7:3 resonance.

Unlike the previous setup (see Sect.~\ref{sec:meth}), the pre-resonance eccentricities of both Ganymede and Callisto would have been smaller than their current values. More precisely, $e_4$ could have ranged between $0$ and $0.007$, while $e_3$ would have been equal to its forced value, which is between $0.0006$ and $0.0009$ (mean value), depending on the chosen value for the eccentricity of Callisto.

We explored this scenario running dedicated numerical simulations: we sampled $e_4$ between $0.001$ and $0.008$ every $0.001$, fixed $e_3$ to its forced value, and took $25$ different initial values of $\lambda_4$ as described in Sect.~\ref{sec:meth}, for a total of $200$ simulations (see \tablename~\ref{tbl3}). For these simulations, we set $k_2/Q=0.01$. In order to obtain the current values of the eccentricities at the end of the evolution, it is necessary that the moons were captured into resonance.

\begin{table}
\caption{Statistics of the outcomes of the resonance crossing for simulations starting with null free eccentricity of Ganymede.}\label{tbl3}
\begin{tabular*}{\tblwidth}{@{}LRRR@{}}
\toprule
initial $e_4$  & no capture  & $e_3$ sub-resonances & other sub-resonances \\ 
\midrule
$0.001$ & $100\%$ & $0\%$ & $0\%$ \\
$0.002$ & $92\%$ & $0\%$ & $8\%$ \\
$0.003$ & $92\%$ & $0\%$ & $8\%$ \\
$0.004$ & $44\%$ & $36\%$ & $20\%$ \\
$0.005$ & $0\%$ & $72\%$ & $28\%$ \\
$0.006$ & $0\%$ & $88\%$ & $12\%$ \\
$0.007$ & $0\%$ & $100\%$ & $0\%$ \\
$0.008$ & $0\%$ & $96\%$ & $4\%$ \\
\bottomrule
\end{tabular*}
\end{table}

None of the simulations were successful. More precisely, in about half of the simulations there was no capture into resonance, so that the final $e_3$ and $e_4$ resulted in too low values; while, in the remaining half, the moons were captured into the 7:3 resonance, but either $e_3$ or $e_4$ increased well above their current values. In \tablename~\ref{tbl3}, we report the statistics of the outcomes for different initial values of $e_4$ (with $e_3$ set to its forced value), distinguishing between capture into sub-resonances involving $e_3$ and sub-resonances not involving it. For values of $e_4\le 0.003$, the most probable outcome is a crossing without capture into resonance, while, for $e_4\ge 0.005$, the probability of capture is $100\%$, with a vast majority of simulations for which the moons were captured into sub-resonances involving $e_3$ when $e_4\ge 0.006$. In particular, the evolution was dominated by the resonance $e_3e_4^3$, which represented the total of captures into $e_3$ sub-resonances for this batch of simulations.

These statistics completely differ from those presented in Sect.~\ref{sec:resu}, where we set a larger initial value of $e_3$. In fact, for initial $e_4\ge 0.006$, we found that the most probable outcome was that the moons crossed the resonance without being captured; here, instead, for similar values of $e_4$, the capture is almost unavoidable. This shows how the eccentricity of Ganymede was determinant in the orbital evolution of the system through the 7:3 resonance. In fact, if its pre-resonance value was smaller, today we would probably observe the Galilean moons' system locked into a four-body resonant chain.

As the evolution of Ganymede's eccentricity depends on the tidal dissipation within the moon, we can attempt to constrain the value of its dissipative parameter $k_2/Q$. The surface of the satellite shows signs of past tectonic activity and large resurfacing, probably due to orbital excitations caused by past resonance crossings (e.g., \citealp{TITTEMORE_1990,MALHOTRA_1991}). The age of the youngest terrains of Ganymede is estimated between 1 and 2 Gyrs ago \citep{ZAHNLE-etal_2003,BABY-etal_2023}. If part of Ganymede's free eccentricity has survived since then, its $k_2/Q$ must be necessarily low. Since we do not know the exact age and the maximum eccentricity reached during the last orbital excitation, we can only estimate that the timescale of eccentricity damping must be at least a few hundred millions of years, so that $k_2/Q\lesssim 0.001$.

In the next subsection, we show some common evolution in the case the two moons are captured into the 7:3 resonance.

\subsection{Sub-resonances involving $e_3$}

Whether the dissipative parameter of Ganymede is set to a small or a large value, capture into the 7:3 resonance generally does not reproduce the current orbital configuration of the system, especially if the eccentricity of Ganymede is directly involved in the resonance. Here, we report two cases for which Ganymede and Callisto are captured into resonance and their eccentricities increase well above their current values of $0.0016$ and $0.0074$, respectively. The first one is obtained starting from a free eccentricity of Ganymede equal to zero (see Sect.~\ref{sec:dis1}), while the second is obtained starting from a free eccentricity slightly larger than today's (see Sect.~\ref{sec:resu}).

\begin{figure*}
\centering
\includegraphics[scale=0.6]{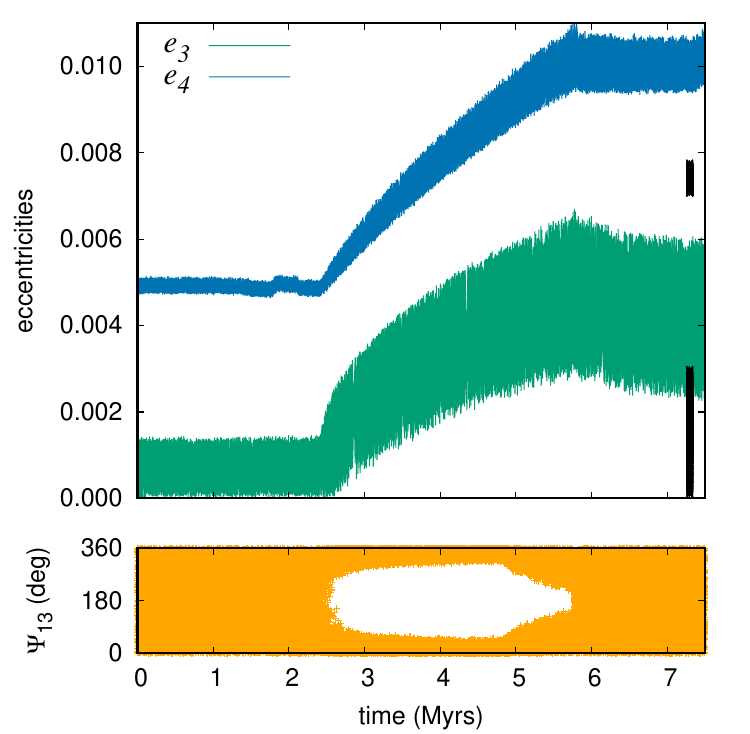}\includegraphics[scale=0.6]{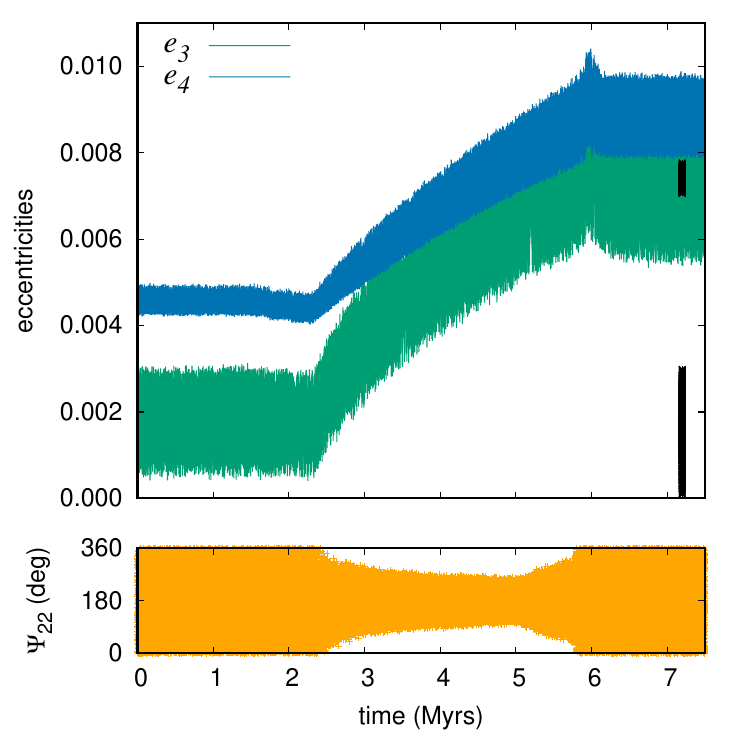}
\caption{Evolution of the eccentricities of Ganymede and Callisto (top) and of the 7:3 resonant argument (bottom), in the case the two moons are captured into 7:3 resonance. On the left, the moons triggered the $e_3 e_4^3$ sub-resonance, while, on the right, the $e_3^2 e_4^2$ sub-resonance.}
\label{fig:e34cap}    
\end{figure*}

The first case is a capture into the $e_3e_4^3$ sub-resonance, which is characterized by the resonant argument $\Psi_{13}=3\lambda_3-7\lambda_4+\varpi_3+3\varpi_4$. As shown in \figurename~\ref{fig:e34cap} (left panel), the increase in the eccentricity of Ganymede is fast, and $e_3$ reaches an average value of $0.0035$ (more than twice its present value) after just a couple of million years. The lock into resonance can persist for many millions of years, allowing the eccentricity of Ganymede to grow to values larger than $0.005$. In the simulation presented in \figurename~\ref{fig:e34cap}, the resonance broke down at about $e_3=0.005$ and $e_4=0.010$ (mean values).

The breaking of the resonance occurred because the resonance widened while the eccentricities increased, so that it overlapped with the nearby sub-resonances. Because of the overall complexity of the resonant region and different dynamical conditions at capture, the disruption of the resonance did not always happen at the same values of the eccentricities. In our simulations, we noted that the 7:3 resonance is more stable and lasts longer when the initial free eccentricity of Ganymede was smaller. This is because the capture into resonance is adiabatic and the system enters the resonance without crossing the separatrix.

The second case is the capture into the $e_3^2e_4^2$ sub-resonance, which is characterized by the resonant argument $\Psi_{22}=3\lambda_3-7\lambda_4+2\varpi_3+2\varpi_4$. As shown in \figurename~\ref{fig:e34cap} (right panel), the increase in Ganymede's eccentricity was even faster than the previous case, with $e_3$ reaching $0.005$ after two million years from the capture. Also for this sub-resonance, the lock into resonance could have persisted for many millions of years. In the simulation presented in \figurename~\ref{fig:e34cap}, the resonance broke down at about $e_3=0.007$ and $e_4=0.009$ (mean values).

Once the two outer moons eventually exited the resonance, their eccentricities could have damped because of tidal dissipation. However, the time interval between the resonance disruption and the current epoch was very short, around two million years. Therefore, we would require very high values of the dissipative parameters of the moons: for Ganymede an extremely large $k_2/Q\approx 0.1$ could be sufficient, but for Callisto no realistic values of its dissipative parameter would have been able to damp its eccentricity on such a short time span.

In the end, the 7:3 resonance had the potentiality to pump the eccentricities of the two outer moons to values much larger than their current ones, especially for Ganymede. This should lead to a revaluation of high-order resonances in the orbital history of moon systems of the solar system. Nevertheless, our results show that the 7:3 resonance is not the source of the observed free eccentricities of Ganymede and Callisto. 

\section{Conclusion}\label{sec:conc}

In this work, we explored the encounter of the 7:3 resonance between Ganymede and Callisto through accurate numerical simulations. Considering the measured tidal dissipation within Io and Jupiter \citep{LAINEY-etal_2009,PARK-etal_2025}, this event should have happened very recently, around two million years ago. We showed that, starting from eccentricities slightly larger than today's, the moons had a probability of about $65\%$ not to be captured into resonance. At the passage through the resonance, Ganymede and Callisto's eccentricities underwent a downward kick; therefore, taking appropriate pre-resonance values of the eccentricities, it was possible to perfectly match the current orbital configuration at the end of the numerical simulations.

Although the probability of capture into the 7:3 resonance was significant, even $100\%$ if the initial free eccentricity of Ganymede was null, such an evolution is generally not compatible with the present-day orbits of the moons, as the eccentricities of Ganymede and Callisto would have increased well above their current values. Temporary capture into particular sub-resonances could have been possible, but the overall probability of this kind of simulations is very low.

From these results, it follows that the tidal dissipation within Ganymede must be low, otherwise its free eccentricity before the resonant encounter would have been completely damped and today Ganymede and Callisto would be almost certainly trapped into the 7:3 resonance. A precise estimation of the amount of tidal heating within the moon would require knowing precisely its orbital history for a time span significantly longer than a few million years. However, given the migration rate and the estimated age of the younger terrains of Ganymede, we can assume that the damping timescale of its eccentricity is at least a few hundred million years, which yields an upper bound for its dissipative parameter $k_2/Q\lesssim 0.001$.

Thanks to the proximity of the 7:3 resonance, with our numerical simulations we managed to accurately reconstruct the orbital evolution of the Galilean moons from the resonance crossing to the present epoch. In doing this, we exploited the tight constraints given by moons' ephemerides, with which we finely tuned initial orbital conditions and revealed important dynamical effects.

In particular, although the orbits of Io and Europa were almost unaffected by crossing the 7:3 resonance between the two outer moons, the Laplace resonance underwent an excitation that pumped the amplitude of its free libration. In a good number of numerical simulations, we obtained a final amplitude close to the current value of $0.061^\circ$.

Since it was first precisely measured \citep{LIESKE_1980}, many researchers have considered the present-day amplitude of libration as the result of a smooth damping from the resonance formation until today. In this way, they provided an estimate of the origin epoch of the Laplace resonance \citep{YODER-PEALE_1981, HENRARD_1983,MALHOTRA_1991}. In this work, we showed that the observed libration amplitude is actually the result of a recent orbital excitation and it has then lost its direct relation with the formation of the three-body resonance between the three inner moons.

This result removes one possible argument in favor of a late tidal capture of the moons into the Laplace resonance \citep{YODER-PEALE_1981,MALHOTRA_1991} versus a primordial origin of the resonance \citep{PEALE-LEE_2002,OGIHARA-IDA_2012,SHIBAIKE-etal_2019,BATYGIN-MORBIDELLI_2020}. However, it is not sufficient to unequivocally discern between the two scenarios and further studies of the past orbital evolution of the moons are necessary. Nevertheless, the results presented in this work could be useful for constraining the dynamical history of the Galilean moons over larger timescales.

Finally, the numerical simulations revealed the effect of an extremely recent dynamical excitation occurred just a few tens of thousands of years ago. This was due to the crossing of a three-body resonance between the three outer Galilean moons, and its main effect was to increase the oscillation amplitude of the eccentricity of Europa. This was the last piece of the evolution that allowed an accurate reconstruction of the Galilean moons' orbital history over the last four million years.

Comparing the billion-year evolution of the Galilean satellites with the length of a day, our study focused on the last minute before midnight, with the crossing of the three-body resonance occurring just at the last second. Although this is a relatively short period of time, we showed that the dynamical evolution of the moon system was very rich and could have followed different pathways, so that today it could have been even possible to admire a four-body resonant chain involving all Galilean moons.\\

\noindent\textbf{Acknowledgments}\\
This work was funded in part by the Italian Space Agency (ASI) through agreement no. 2022-16-HH.0.

\bibliographystyle{cas-model2-names}

\bibliography{freelib}

\end{document}